\nonstopmode
\documentclass[aip,twocolumn]{revtex4}

\usepackage{amsmath,amssymb,graphicx}
\usepackage{graphicx}
\usepackage{mdwmath}
\usepackage{mdwtab}

\begin{document}

\def\clr{\color{red}}
\def\jlt{ J.\ Lightwave\ Technol.\ }
\def\ptl{ IEEE Photon.\ Tech.\ Lett.\ }
\def\omex{ Opt.\ Mat.\ Express }
\def\pr{Phys.\ Rev. }

\def\gammat{\tilde\gamma}
\def\etat{\tilde\eta}
\def\Pt{\tilde P}
\def\phit{\tilde\phi}
\def\LP{\rm LP}
\def\NLSE {nonlinear Schr\"{o}dinger equation}

\def\opex{ Opt.\ Express }
\def\ao{ Appl.\  Opt.\ }
\def\ap{ Appl.\  Phys.\ }
\def\apa{ Appl.\  Phys.\ A }
\def\apb{ Appl.\  Phys.\ B }
\def\apl{ Appl.\ Phys.\ Lett.\ }
\def\apj{ Astrophys.\ J.\ }
\def\bell{ Bell Syst.\ Tech.\ J.\ }
\def\jqe{ IEEE J.\ Quantum Electron.\ }
\def\jlt{ J.\ Lightwave\ Technol.\ }
\def\ptl{ IEEE Photon.\ Tech.\ Lett.\ } 
\def\assp{ IEEE Trans.\ Acoust.\ Speech Signal Process.\ }
\def\aprop{ IEEE Trans.\ Antennas Propag.\ }
\def\mtt{ IEEE Trans.\ Microwave Theory Tech.\ }
\def\iovs{ Invest.\ Ophthalmol.\ Vis.\ Sci.\ }
\def\jcp{ J.\ Chem.\ Phys.\ }
\def\jmo{ J.\ Mod.\ Opt.\ }
\def\josa{ J.\ Opt.\ Soc.\ Am.\ }
\def\josaa{ J.\ Opt.\ Soc.\ Am.\ A }
\def\josab{ J.\ Opt.\ Soc.\ Am.\ B }
\def\jpp{ J.\ Phys.\ (Paris) }
\def\nat{ Nature (London) }
\def\oc{ Opt.\ Commun.\ }
\def\ol{ Opt.\ Lett.\ }
\def\pl{ Phys.\ Lett.\ }
\def\pra{ Phys.\ Rev.\ A }
\def\prb{ Phys.\ Rev.\ B }
\def\prc{ Phys.\ Rev.\ C }
\def\prd{ Phys.\ Rev.\ D }
\def\pre{ Phys.\ Rev.\ E }
\def\prl{ Phys.\ Rev.\ Lett.\ }
\def\rmp{ Rev.\ Mod.\ Phys.\ }
\def\pspie{ Proc.\ SPIE\ }
\def\sjqe{ Sov.\ J.\ Quantum Electron.\ }
\def\vr{ Vision Res.\ }

\def\cleo{ {\it Conference on Lasers and Electro-Optics }}
\def\assl{ {\it Advanced Solid State Lasers }}
\def\tops{ Trends in Optics and Photonics }

\title{Nonlinear switching in a concentric ring core chalcogenide glass optical fiber for passively mode-locking a fiber laser}

\author{Elham Nazemosadat}
\affiliation{Department of Electrical Engineering and Computer Science, University of Wisconsin-Milwaukee, Milwaukee, WI 53211, USA.}

\author{Arash Mafi}
\affiliation{Department of Electrical Engineering and Computer Science, University of Wisconsin-Milwaukee, Milwaukee, WI 53211, USA.}
\thanks{corresponding author.}

\begin{abstract}
We propose an all-fiber mode-locking device 
which operates based on nonlinear switching
in a novel concentric ring core fiber structure.
The design is particularly attractive given the ease of fabrication and 
coupling to other components in a mode-locked fiber laser cavity. 
The nonlinear switching in this coupler is studied and the relative power transmission is obtained. The analysis shows that
this nonlinear switch is practical for mode-locking fiber lasers and is forgiving to fabrication errors.
\end{abstract}
\maketitle

\noindent Optical pulse formation in passively mode-locked lasers is achieved by employing a nonlinear 
optical device, referred to as a saturable absorber (SA). 
These devices provide intensity discrimination inside the laser cavity,
which leads to the suppression of continuous-wave lasing 
and generation of optical pulse trains. 
Common fiber-integrable SAs have deficiencies such as long term reliability, 
thermal issues, and sensitivity to environmental perturbations, 
which limit their usefulness for mode-locked fiber laser applications.
Also, common SAs have become a limiting factor in boosting up the pulse 
energy and peak power of mode-locked lasers to higher values.
Hence, there is a strong need to develop robust alternative mode-locking techniques.

Recently nonlinear mode-coupling in waveguide arrays 
and multi-core fibers~\cite{Winful, Proctor, Hudson, Chao} and also nonlinear multi-modal interference
in graded-index multimode fibers~\cite{MMF} have been introduced as 
alternative techniques that address most of the limitations faced by common SAs.
In Ref.~\cite{MCFvsMMF} we compared the nonlinear switching behavior in multi-core versus multi-mode fibers
and showed that a much higher switching power is required in the latter case,
due to the large difference in the propagation constant of the modes. This feature makes multi-mode fiber couplers potential mode-locking devices for future high peak power lasers. 
For lower peak powers, as is the case in most present day mode-locked fiber lasers, it is preferred to use multi-core couplers.

Our recent study on multi-core fiber array couplers 
showed that going from a two-core fiber geometry to a higher number of cores
does not improve the nonlinear switching performance of the device considerably, if any at all~\cite{SA}. 
Hence, according to our studies, a
two-core fiber coupler with optimized parameters is the optimum solution to achieve an all-fiber mode-locking device.

\begin{figure}[htb]
\centering
\includegraphics[width=1in]{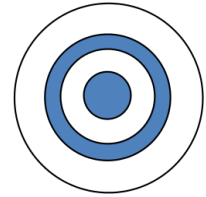}
\caption{Cross sectional view of the fiber structure. The cores
are identified with a darker shade.}
\label{fig:fiber}
\end{figure}

In this paper, we propose a novel concentric ring core (CRC) fiber geometry as a mode-locking device. This fiber has two concentric cores 
composed of a circular core located at the center of the fiber and a ring core placed around the central core, as shown in Fig.~\ref{fig:fiber}. 
The main reason for choosing this geometry is its simple fabrication process. Given that conventional fibers are fabricated radially, maintaining 
the sensitive specifications of this fiber during the drawing process is easier than non-concentric conventional two-core fibers.
Another related reason is that
aligning the central core of this fiber with the single mode fibers (SMFs) in the laser cavity is much easier and more robust compared with 
conventional two-core geometries, especially in the presence of the inevitable variations in the fiber geometry due to the fabrication and drawing process.
   
In this Letter, the parameters of this two-core fiber mode-locking device are designed and 
the nonlinear propagation of a pulse in this coupler is analyzed. 
To lower the required switching power,
it would be preferred to build the coupler from a fiber with a higher nonlinearity compared to silica~\cite{Zakery}.
Chalcogenide glass is a suitable candidate because it has a nonlinear coefficient up to 1000 times larger than silica~\cite{Lenz,Sugimoto}.  
																																																																																																																																																																																																																																																																																																																																																																																																																																																																																																																																																																																																																																																																																																																																																																																																																																																																																																																																																																																																																																																																																																																																																																																																																																																																																																																																																																																																																																																																																																																																																																																																																																																																																																																																				
\begin{figure}[htb]
\centering
\includegraphics[width=2.5in]{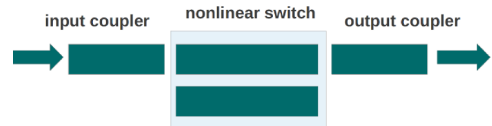}
\caption{An all-fiber nonlinear switching device composed of a CRC coupler placed in between two SMFs.}
\label{fig:coupler}
\end{figure}

The all-fiber mode-locking device is shown in Fig.~\ref{fig:coupler}, where the beam is injected into the CRC fiber coupler using an input
SMF and is collected at the output from another SMF. 
The injected beam from the input 
SMF is in the form of an azimuthal symmetric nearly Gaussian beam, so by centrally aligning the core of the input SMF with the central core of the CRC coupler, 
only zero angular momentum modes will be excited in the coupler~\cite{MafiMMI1}.
For low input optical powers launched into the nonlinear coupler,
optical power is periodically exchanged among the two cores because of their modal overlap.  
This linear coupling is most efficient when the propagating modes have identical propagation constants~\cite{Saleh}. 
Therefore, to achieve maximum linear coupling, the size of each core and the separation among them are designed such that the propagation constant of the two propagating modes are identical.
For high input optical powers, nonlinear effects arise and alter the refractive index of each waveguide. 
This will detune the effective propagation constants of the modes and will consequently reduce 
the power exchange efficiency between the cores, which means light will be mainly preserved in the launch core~\cite{Jensen}. 
Hence, if an optical pulse is transmitted through a fiber coupler with a length equal to the coupler's half-beat-length,
its low intensity tails will be efficiently ``linearly'' coupled to the adjacent core, 
while its high intensity center peak will remain in the launch core~\cite{Friberg}. 
If only the light in the launch core is collected at the output,
the required power-dependent transmission for mode-locking is obtained~\cite{Winful,Proctor}. 

\begin{figure}[htb]
\centering
\includegraphics[width=2.5in]{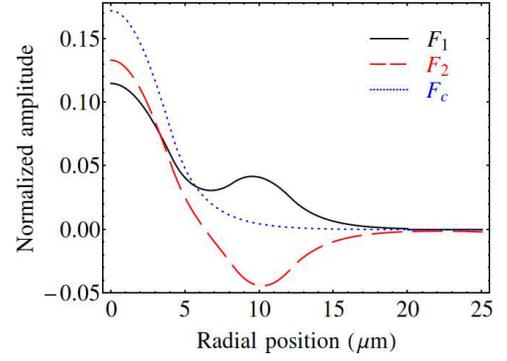}
\caption{The normalized spatial mode profile of the first order (solid black), 
second order (dashed red) and center core (dotted blue) modes of the CRC fiber are displayed.}
\label{fig:mode-profile}
\end{figure}

The fiber profile is designed such that it supports only two zero-angular momentum guided modes referred to as the first and second order guided supermodes.
The electric field in the CRC can be expressed as
the summation of these guided modes as  
\begin{align}
\label{eq:ElectricalFieldEO}
E(x,y,z,t)=\sum_{\mu=1,2}A_{\mu}(z,t)F_{\mu}(x,y)e^{i(\omega_{0}t-\beta_{0,\mu}z)},
\end{align}
where $A_{\mu}(z,t)$ is the slowly varying envelope of the electric field of the $\mu$th mode with the normalized spatial 
distribution $F_{\mu}(x,y)$ and propagation constant $\beta_{0,\mu}$.
The generalized nonlinear Schr\"{o}dinger equation (GNLSE) describing the longitudinal evolution of 
the pulse in this fiber coupler can be written as~\cite{Poletti,Mafi}
\begin{align}
\label{eq:master}
\nonumber
\dfrac{\partial A_{\mu}}{\partial z}&=
i~\delta \beta_{0,\mu}A_{\mu}-\delta \beta_{1,\mu}\dfrac{\partial A_{\mu}}{\partial t}
-i\dfrac{\beta_{2,\mu}}{2}\dfrac{\partial^2 A_{\mu}}{\partial t^2}\\
&+i(\dfrac{n_2\omega_0}{c})\sum_{\nu,\kappa,\rho=1,2}f_{\mu\nu\kappa\rho}A_\nu A_\kappa A^\ast_\rho,
\qquad \mu=1,2,
\end{align}
The indices can take the value of $1$ or $2$ corresponding to first and second order modes. $\beta_{1,\mu}$ is the group delay per unit length, and
$\beta_{2,\mu}$ is the group velocity dispersion (GVD) parameter of the $\mu$th mode.
We define
$\delta\beta_{0,1}=-\delta\beta_{0,2}=\beta_{0,1}-\beta_{0,{\rm ref}}$ and 
$\delta\beta_{1,1}=-\delta\beta_{1,2}=\beta_{1,1}-\beta_{1,{\rm ref}}$, where 
$\beta_{i,{\rm ref}}=(\beta_{i,1}+\beta_{i,2})/2$ for i=~0,1. $n_2$ is the nonlinear 
index coefficient and $\omega_0$ is the carrier frequency.
The nonlinear coupling coefficients are given by
\begin{align}
\label{eq:nonlinear_coefficient}
f_{\mu\nu\kappa\rho}=\iint{dx dy} F^\ast_\mu F_\nu F_\kappa F^\ast_\rho,
\end{align}
where the mutually orthogonal spatial profiles are assumed to be normalized according to
$\iint F^2_1 dxdy=\iint F^2_2 dxdy=1$. The spatial mode profiles of the supermodes of the CRC coupler, $F_{\rm 1}$ and $F_{\rm 2}$, and also that of the center core, $F_{\rm c}$, 
were found using the finite element method~\cite{Lenahan} and are shown in
Fig.~\ref{fig:mode-profile}. $F_{\rm c}$ is calculated for the central core in the absence
of the outer ring.
The beam is launched into and collected from the 
center core of the CRC fiber coupler.
The transmittance is defined as the energy at the output of the center core divided by the total energy injected into the coupler, given by
\begin{equation}
\tau=\dfrac{\int_{-\infty}^{+\infty} |A_c(L_c,t)|^2 dt}{\int_{-\infty}^{+\infty} (|A_1(0,t)|^2+|A_2
(0,t)|^2)dt},
\label{eq:tauSMFGIMFSMF}
\end{equation}
where $A_{c}$ is the field amplitude of the center core, defined as 
\begin{align}
\nonumber
A_c(z,t)&=A_1(z,t)\iint F^\ast_cF_1 dxdy\\
&+A_2(z,t)\iint F^\ast_cF_2 dxdy,
\label{eq:core-field-amplitude}
\end{align}

The refractive index of both cores is $n_{co}=2.8035$, while that of the cladding is $n_{cl}=2.8$. 
The center core has a radius of $r_{1}=4.1~{\rm \mu m} $, the inner radius of the ring
is $r_{2}=8.2~{\rm \mu m}$ while its width is $w_{2}=3.94~{\rm \mu m}$. 
The nonlinear refractive index is $n_2=0.9\times10^{-17}~{\rm m^2/W}$ and the operating wavelength is considered $\lambda_0=1550$~nm~\cite{Fu}.
With these parameters, each individual core supports only one zero angular momentum mode.
It should be pointed out that in order to minimize the 
splicing loss between the cavity SMFs and the CRC,
the dimensions and numerical aperture of the center core are similar to that of the
two SMFs. 
The length of the coupler is $1.75$~mm, which is equal to the half-beat-length of the coupler, 
given by $L_c=\pi/(2C)$, where $C=\delta\beta_{0,1}$ is the linear coupling coefficient among 
the two modes~\cite{Saleh}. It should be noted that at the considered wavelength, the particular 
chalcogenide glass that we have considered here has a negative dispersion parameter equal to $D=-504~$ps/nm/km~\cite{Fu},
and thus operates in the normal dispersion regime.

\begin{figure}[htb]
\centering
\includegraphics[width=2.5in]{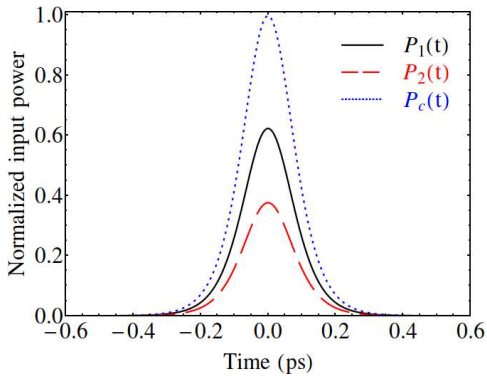}
\caption{The normalized excited power of the first (solid black) and second (dashed red) order supermodes and the normalized power launched in the center core (dotted blue) at the input facet of the CRC fiber.}
\label{fig:mode-power}
\end{figure}

\begin{figure}[htb]
\centering
\includegraphics[width=2.5in]{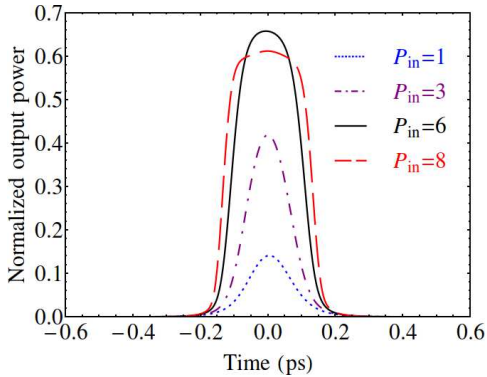}
\caption{The normalized pulse power at the output of the center core for different input peak powers (units are KW) when the pulse width is $t_0=0.1~ps$. For low
input powers, most of the pulse energy is transferred to the ring core, whereas for higher input powers the majority of the pulse energy remains in the center core.
}
\label{fig:Pout-center}
\end{figure}

We assume that a hyperbolic secant optical pulse with a pulse width of $t_0$ is launched 
into the nonlinear coupler through the input SMF; the nearly Gaussian spatial mode profile 
of the input beam with a temporal power profile of $P_c(t)$ excites the supermodes of the 
CRC fiber with power profiles $P_{1}(t)$ and $P_{2}(t)$ with
normalized peak powers $p_{1}$ and $p_{2}$, respectively, where it can be shown $p_{1}+p_{2}\approx1$. 
The normalized peak excitation power of each supermode at the CRC input facet is 
calculated using the overlap integral of the input beam and that particular supermode, 
where in this case is found to be $p_{1}=62.3\%$ and $p_{2}=37.6\%$. 
While Fig.~\ref{fig:mode-power} displays the normalized power profiles of the supermodes and also that of the central core at the CRC fiber input facet, Fig.~\ref{fig:Pout-center} shows how the pulse shape will look like at the output of the CRC as a function of the input pump power.
                                                                                                                                                                           
We have applied split step Fourier method~\cite{AgrawalBook} to solve Eq.~\ref{eq:master} and analyze the propagation of the
pulse inside the nonlinear coupler.
For low powers, the CRC fiber operates similar to a linear directional coupler at half-beat-length and most of the pulse energy is 
transferred to the ring core.
For intense pulses, the central portion of the pulse remains in the central waveguide while 
the low intensity wings of the pulse are transferred to the ring core. 
In Fig.~\ref{fig:Pout-center} 
the output of the center core is shown for different input peak powers, for the case where the input pulse has a width of $t_0=0.1~\rm {ps}$. 
As the input peak power is increased, nonlinear terms also become larger and
because of the large amount of the induced frequency
chirp imposed on the pulse due to self-phase modulation, even weak dispersive effects lead to significant pulse shaping~\cite{AgrawalBook}.
This feature can be observed in Fig.~\ref{fig:Pout-center}, where for an input peak power of $P_{in}=8~\rm {KW}$ 
the output pulse will be in the form of a rectangular-shaped pulse with relatively sharp leading and trailing edges.  

\begin{figure}[htb]
\centering
\includegraphics[width=2.8in]{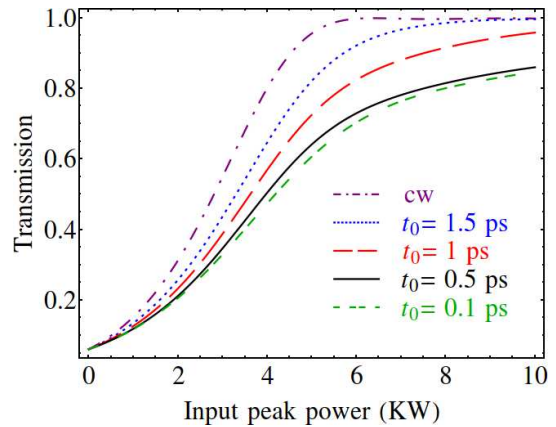}
\caption{The relative power transmission is plotted as a function of the input peak power for different pulse widths.}
\label{fig:Tau-different-t0}
\end{figure}

Fig.~\ref{fig:Tau-different-t0} displays the relative power transmission as a function of the input peak power for various input pulse 
durations. It can be observed that a higher input power is required to switch the output from 
$\tau_{min}$ to $\tau_{max}$ as the duration of the pulse gets shorter. Also the maximum transmission decreases as the pulse duration is shortened; 
for the continuous wave (CW) input $\tau_{max}=1$, while it drops to $\tau_{max}\approx0.8$ for $t_0=0.1~ps$.
The reason is that the dispersion effects become more 
significant for short pulse widths, and the resultant pulse broadening prevents complete power transfer.

\begin{figure}[htb]
\centering
\includegraphics[width=2.5in]{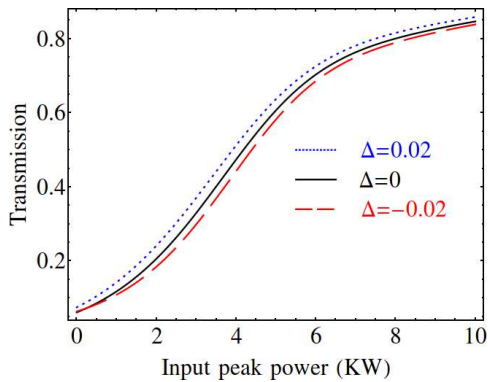}
\caption{The relative power transmission is plotted as a function of the input peak power for no changes (solid black), $2\%$ increase (dashed red) 
and $2\%$ decrease (dotted red) in the size of the core radii when the pulse width is $t_0=0.1~ps$}
\label{fig:Tcomparison}
\end{figure}

During the fabrication process, random core diameter fluctuations will be inevitable; to estimate the effect of these possible variations on the coupler's transmission,
we altered the core sizes and calculated the relative power transmission.
$r'_{1}$ and $w'_{2}$ are defined as the radius and width of the center and ring core after considering the size changes, respectively, 
and $\Delta=(r'_{1}-r_1)/r_1=(w'_{2}-w_2)/w_2=\pm2\%$ is the
amount of the considered core size variations.
The obtained results which are displayed in Fig.~\ref{fig:Tcomparison} 
show that possible variations in the core sizes do not affect the transmittance of the nonlinear coupler significantly, 
making the proposed configuration a robust and stable mode-locking element.

In summary, we have proposed a novel concentric ring core fiber structure made of chalcogenide glass to be used 
as an all-fiber mode-locking element. The nonlinear switching behavior of this coupler is 
analyzed using the supermodes of the waveguide. The chalcogenide glass has a large chromatic dispersion.
In order to avoid substantial pulse broadening due to dispersion, the length of the coupler (half-beat-length) 
should be reduced as much as possible. This in turn can be achieved by reducing the separation between the 
central core and the ring core and increasing the coupling between cores. On the other hand, increasing the 
coupling between the cores increases the switching power, which may not be desirable. Therefore, the optimal 
design is a compromise between the pulse broadening and switching power. The pulse broadening and switching power 
can also be modified by changing the geometrical parameters and glass compositions. The switching power is slightly over 
4~KW in the present structure.

The authors acknowledge support from the Air Force Office of Scientific Research under Grant FA9550-12-1-0329.


\begin{thebibliography}{99}
\bibitem{Winful}
H. G. Winful, and D. T. Walton, \ol {\bf 17}, 1688 (1992).
\bibitem{Proctor}
J. L. Proctor, and J. N. Kutz, \ol {\bf 30}, 2013 (2005).
\bibitem{Hudson}
D. D. Hudson, K. Shish; T. R. Schibli, J. N. Kutz, D. N. Christodoulides, R. Morandotti, and S. T. Cundiff, 
\ol {\bf 33}, 1440 (2008).
\bibitem{Chao}
Q. Chao, D. D. Hudson, J. N. Kutz, and S. T. Cundiff, IEEE Photon. J. {\bf 4}, 1438 (2012).
\bibitem{MMF}
E. Nazemosadat and A. Mafi, \josab {\bf 30}, 1357 (2013).
\bibitem{MCFvsMMF}
E. Nazemosadat and A. Mafi, \opex {\bf 21}, 30739 (2013).
\bibitem{SA}
E. Nazemosadat and A. Mafi, \josab {\bf 30} 2787 (2013).
\bibitem{Zakery}
A. Zakery and M. Hatami,
J. Phys. D: Appl. Phys. {\bf 40} 1010 (2007).
\bibitem{Lenz}
G. Lenz, J. Zimmermann, T. Katsuf uji, M. E. Lines, H. Y. Hwang, S. Sp\"{a}lter, R. E. Slusher, 
S.-W. Cheong, J. S. Sanghera and I. D. Aggarwal, \ol {\bf 25}, 254 (2000).
\bibitem{Sugimoto}
N. Sugimoto, H. Kanbara, S. Fujiwara, K. Tanaka, Y. Shimizugawa, and K. Hirao,
\josab {\bf 16}, 1904 (1999).
\bibitem{MafiMMI1}
A. Mafi, P. Hofmann, C. Salvin, and A. Sch\"{u}lzgen, \ol {\bf 36}, 3596 (2011).
\bibitem{Saleh}
B. E. A. Saleh, and M. C. Teich, Fundamentals of Photonics, 2nd ed. (Wiley, 2007).
\bibitem{Jensen}
S. M. Jensen, IEEE J. Quantum Electron. {\bf 18}, 1580 (1982).
\bibitem{Friberg}
S. R. Friberg, A. M. Weiner, Y. Silberberg, B. G. Sfez, and P. S. Smith, \ol {\bf 13}, 904 (1988).
\bibitem{Poletti}
F. Poletti and P. Horak, \josab {\bf 25}, 1645 (2008).
\bibitem{Mafi}
A. Mafi, \jlt {\bf 30}, 2803 (2012).
\bibitem{Lenahan}
T. A. Lenahan, Bell\ Syst.\ Tech.\ J.\ {\bf 62}, 2663-2694 (1983).
\bibitem{Fu}
L. Fu, M. Rochette, V. Ta'eed, D. Moss, and B. Eggleton, \opex {\bf 13}, 7637 (2005).
\bibitem{AgrawalBook}
G. P. Agrawal, Nonlinear fiber optics, 5th ed. (Elsevier Science, 2013).
\end{thebibliography}
\end{document}